\documentclass{article}
\usepackage{amsmath,amssymb,amsfonts}

\newtheorem{theorem}{Theorem}[section]
\newtheorem{lemma}[theorem]{Lemma}

\title{Initial data for a head on collision of two
  Kerr-like black holes with close limit}
\author{Sergio Dain\\
  Max-Planck-Institut f\"ur Gravitationsphysik\\
  Am M\"uhlenberg 1\\
  14476 Golm\\
  Germany}

\begin{document}
\maketitle

\begin{abstract}
  We prove the existence of a family of initial data for the Einstein
  vacuum equation which can be interpreted as the data for two
  Kerr-like black holes in arbitrary location and with spin in
  arbitrary direction. This family of initial data has the following
  properties: (i) When the mass parameter of one of them is zero or
  when the distance  between them goes to infinity, it reduces
  exactly to the Kerr initial data.  (ii) When the distance between
  them is zero, we obtain exactly a Kerr initial data with mass and
  angular momentum equal to the sum of the mass and angular momentum
  parameters of each of them. The initial data depends smoothly on the
  distance, the mass and the angular momentum parameters.

\vspace{0.5cm}

\end{abstract}

\section{Introduction}

Perturbation analysis has been used successfully  in the study of
black-hole collisions (see \cite{Pullin99} and references therein and
also \cite{Baker00} for a recent approach which combines full numerical
calculations with perturbation theory).
In the case of a  head-on collision a pure linear analysis provide a
remarkable accurate description of the late state of the
collision\cite{Pullin99}. 
This calculation is based on the fact that the head-on initial data 
\cite{Misner60}\cite{Brill63} can be written as a perturbation of the
Schwarzschild initial data, the separation between the black-holes is
the perturbation parameter.  When the black holes have
intrinsic angular momentum, the natural generalization of this idea
is to consider   perturbations  around a Kerr background. The problem
now is
to  construct proper initial data:   a solution of the vacuum 
constraints equations that represents two black holes; such that, when
the separation between them is zero  we recover the initial data  of
the final  Kerr black hole.  This  property is 
usually called the `close limit' of the initial data. 
The standard Bowen-York data\cite{York} do not have a close limit to Kerr,
since it does not include the Kerr initial data for any choice of the
parameters. In \cite{Price98} and  \cite{Price99} it has been computed
numerically a class of axially symmetric initial data that has the
desired close limit to the Kerr, and  it has also  been  studied
the linear evolution of them. However this initial data do not have the
topology of two black holes, an extra undesirable  singularity
appear in the solution. 
The purpose of this article is to overcome this problem, 
generalize the construction for non-axially symmetric data and  give
a rigorous existence proof for this class of initial data.

The plan of the paper is the following. In section \ref{constraints}
we give a brief summary of the standard conformal technique for
solving the constraints equations with many asymptotically flat ends.
This class of initial data contain, in general, apparent horizons
around the ends. The existence of apparent horizons lead us to
interpret these data as representing initial data for black-holes
collision. The evolution of them will presumably contain an
event horizon, the final picture of the whole space time will be
similar to the one shown in figure 60 of \cite{Hawking73}, which
represents a collision and merging of two black holes. We describe in
this section a set of existence theorems that are appropriate for our
present purpose\cite{Dain99}. In section \ref{Kerrini} we summarize
properties of the Kerr initial data that have been proved in \cite{Dain00c}.
Together with the existence theorems of section \ref{constraints}
these are the essential tools for our construction. The new initial
data is given in sections \ref{sk} and \ref{kk}. The idea is very
similar to the one given in \cite{Dain00c}, the important difference
is the choice of the conformal metric and the conformal extrinsic
curvature.  In section \ref{sk} we construct  remarkably simple
initial data that represent a Schwarzschild and a Kerr black hole,
with a close limit to a final Kerr black hole. In section \ref{kk} we
generalize to two Kerr black holes at arbitrary location and with spin
in arbitrary directions. The Kerr metric is characterized by the mass
$m$ and the angular momentum per mass $a$.  The initial data given in
section \ref{kk} depend smoothly on the parameters $m_1, J^a_1$,
$m_2, J^a_2$ and $L$, where $m_1$ and $m_2$ are the mass parameter of
each of the black holes, $J^a_1$ and $J^a_2$ are the spin vectors of
each of them and $L$ is the distance between the black holes. We
define $a^2_1=m_1^2|J_1|^2$ and $a^2_2=m_2^2|J_2|^2$, we always chose
$m_1^2>a^2_1$ and $m_2^2>a^2_2$.  The data satisfies the following two
properties:
\begin{itemize}
\item[(i)] \emph{Far limit:} When $m_1$    is zero, we obtain the Kerr 
initial data with mass $m_2$  and angular momentum $J_2$. If we fix
the location   of  the hole $1$ and take the limit
$L=\infty$, we obtain  the Kerr initial data with
mass $m_1$ and angular momentum  $J_1$. The same is true if we exchange the 
holes $1$ and  $2$, because  the data is symmetric in the 
parameters $1$ and $2$. 

\item[(ii)] \emph{Close limit:}  When $L=0$ we obtain a Kerr initial data
with mass $m_1+m_2$ and angular momentum $J_K=J_1+J_2$.
\end{itemize}

In section \ref{fc} possible generalizations are
discussed. Finally, in appendix \ref{ap}, we state some known
results about the momentum constraint in  axial symmetry in a coordinate
independent  way.

\section{Solutions of the vacuum constraint equations with many
  asymptotically flat ends}\label{constraints}

The  conformal approach to find solutions of the constraint
equations with many asymptotically flat end points $i_k$ is the
following (cf. \cite{Choquet99}, \cite{Choquet80} and the reference
given there.  The setting outlined here, where we have to solve
(\ref{diver}), (\ref{Lich})  on the compact manifold  has been studied in
\cite{Beig}, \cite{Friedrich88},
\cite{Friedrich98}, see also \cite{Beig94},  \cite{Husa98} for an
interesting application of this formalism). Let $S$ be a compact manifold, denote by $i_k$ a finite number of points in $S$, and define the
manifold $\tilde S$ by 
$\tilde S = S \setminus \bigcup i_k$. We assume that
$h_{ab}$ is a positive definite metric on $S$, with covariant derivative
$D_a$, and 
$\Psi^{ab}$ is  a trace-free symmetric tensor, which satisfies   
\begin{equation} 
\label{diver}
D_a \Psi^{ab}=0 \quad\mbox{on}\quad \tilde S.
\end{equation}
Let $\theta$ a solution of 
\begin{equation} 
\label{Lich}
L_h \theta=-\frac{1}{8}\Psi_{ab}\Psi^{ab}\theta^{-7}
\quad \mbox{on}\quad \tilde S , 
\end{equation}
where   $L_h=D^aD_a-R/8$ and $R$ is the Ricci scalar of the metric $h_{ab}$.
Then the physical fields  $(\tilde h, \tilde
\Psi)$ defined by  $\tilde{h}_{ab} =
\theta^4 h_{ab}$ and $\tilde{\Psi}^{ab} = \theta^{-10}\Psi^{ab}$ will
satisfy the vacuum constraint equations on $\tilde S$. To ensure
asymptotic flatness of the data at the points $i_k$ we require at
each point $i_k$
\begin{equation} 
\label{Psii}
\Psi^{ab} = O(r^{-4}) \quad\mbox{as}\quad r \rightarrow 0,  
\end{equation}
\begin{equation} 
\label{thetai}
\lim_{r\rightarrow 0} r\theta = c_k,
\end{equation}
where the $c_k$ are positive constants,  $x^i$ are normal coordinates
centered at $i_k$ and   $r = (\sum_{i=1}^3 (x^j)^2 )^{1/2}$. 

In order to get existence of solutions for equations (\ref{diver}) and
(\ref{Lich}) with the boundary conditions (\ref{Psii}) and
(\ref{thetai}), we have to impose some restrictions  on the conformal
metric  $h_{ab}$.  We require  that the Ricci scalar  of
$h_{ab}$ is positive
\begin{equation}
  \label{eq:posR}
  R>0.
\end{equation}
It is physically reasonable to impose that  $h_{ab}$ is smooth on
$\tilde S$. However, the differentiability of   $h_{ab}$ at the ends
$i_k$ is a very delicate issue, since it characterizes the fall-off
behavior of the initial data near space
like infinity. 
Strong assumptions (for example smoothness) can rule out physical
important initial data (for example the Kerr initial data). A suitable condition for $h_{ab}$,  which in
particular include the Kerr initial data as we will see, is the following 
\begin{equation}
  \label{eq:sobh}
  h_{ab} \in W^{4,p}(S), \quad p\geq 3/2.
\end{equation}
(see e.g. \cite{Adams} for the
definitions of the Sobolev, Lebesgue and H\"older spaces  $W^{s,p}$,
$L^p$ and $C^{m,\alpha}$, $0<\alpha<1$). In particular this
condition  implies that $h_{ab}$ is $C^{2,\alpha}(S)$. 
Existence of solutions
for the constraint equations under the assumption  
(\ref{eq:posR})  and (\ref{eq:sobh})
were proved in  \cite{Dain99}, in the following we briefly summarize
these results. We will present simplified versions of the theorems,
which are sufficient for our present purpose. 

In order to find solutions of equation (\ref{Lich}) with the boundary
condition (\ref{thetai}) we look first for  functions  $\theta_{i_k}$ which
satisfies the linear equation
\begin{equation}
  \label{eq:Green}
L_{h} \theta_{i_k}=0, \text{ in } S \setminus \{i_k\},  
\end{equation}
and at $i$
\begin{equation}
  \label{eq:bGreen}
\lim_{r\rightarrow 0} r  \theta_{i_k}=1, 
\end{equation}
where $i_k\in S$ is an arbitrary
point.  Denote by $B_{i_k}(\epsilon)$ the open ball with
center $i_{k}$ and radius $\epsilon > 0$, where $\epsilon$ is chosen small enough such
that $B_{i_k}$ is a convex normal neighborhood of $i_{k}$. Choose a
function $\chi_{i_k} \in C^\infty (S)$  which is non-negative and
such that $\chi_{i_k} = 1$ in $B_{i_k}(\epsilon/2)$ and $\chi_{i_k} = 0$ in 
$S \setminus B_{i_k}(\epsilon)$. We have the following lemma.
\begin{lemma} 
\label{existenceGreen}
Assume that $h_{ab}$ satisfies  (\ref{eq:posR})  and  (\ref{eq:sobh}). Let
$i_k\in S$ an arbitrary point. Then, there exists a unique
solution $\theta_{i_k}$ of  equation (\ref{eq:Green}) which satisfies
condition (\ref{eq:bGreen}) at $i_k$. Moreover,  
$\theta_{i_k}>0$ in $S\setminus \{i_k\}$ and we can write
$\theta_{i_k}=\chi_{i_k}/r+g_{i_k}$, with $g_{i_k}\in C^{1,\alpha}(S)$.  
\end{lemma}

For each of the end points $i_k$ we have a corresponding function
$\theta_{i_k}$, we define $\theta_0$ by
\begin{equation}
  \label{eq:sumgreen}
\theta_0= \sum_{k=0}^n c_k \theta_{i_k}.  
\end{equation}
The initial data will have $n+1$ asymptotic ends, the relevant cases
for us are $n=0,1,2$. It is important to note that $\theta_0^{-1}$ is
in $C^\alpha(S)$, it is non-negative and vanishes only at the points
$i_k$.

To obtain the solution $\theta$, we write $\theta = \theta_0 + u$ and
solve on $S$ the following equation for $u$
\begin{equation} 
\label{gouu}
L_h u
=-\frac{1}{8} \theta_0^{-7} \Psi_{ab}\Psi^{ab} (1+\theta_0^{-1}u)^{-7}.
\end{equation}
We have the following existence result.
\begin{theorem} 
\label{Beig}
Assume the metric $h_{ab}$ satisfies   (\ref{eq:posR}) and  (\ref{eq:sobh}),
and that $\theta_0^{-7}\Psi_{ab}\Psi^{ab}\in L^q(S)$, $q\geq 2$. 
Then, there exists a unique non-negative solution $u\in W^{2,q}(S)$  
of equation (\ref{gouu}). We have $u>0$ unless $\Psi_{ab}\Psi^{ab}=0$.
\end{theorem}
We note that our assumptions on $\Psi^{ab}$ impose rather mild
restrictions, which are, in particular, compatible with the fall off
requirement (\ref{Psii}).

We turn now to  the linear  equation
(\ref{diver}). A smooth solution on $S$ can easily be obtained by known
techniques (cf.   \cite{Choquet99}) However,  in that case the initial data will have vanishing momentum and
angular momentum. To obtain data without this restriction, we have to
consider fields $\Psi^{ab} \in C^{\infty}(\tilde{S})$ which are singular at
$i_k$. The existence of initial data with non trivial momentum and
angular momentum has been studied in \cite{Beig96}, where the role of
the conformal symmetries is also discussed. In \cite{Dain99} we  have 
generalized some of the results proved  in
\cite{Beig96} for smooth metrics   to  metrics in the class
(\ref{eq:sobh}). We will present here a simplified version of these
results. Consider the following tensor in $\mathbb{R}^3 \setminus \{0\}$
\begin{equation}
 \label{eq:PsiJp}
\Psi_J^{ab} =\frac{3}{r^3}(n^a  \epsilon^{bcd} J_c n_d +  
n^b\epsilon^{acd} J_c n_d),
\end{equation}
where $J^a$ is a constant vector and $n^a=x^a/r$. This tensor is trace-free and
divergence free  with respect to the flat 
metric.  The vector $J^a$ will give the angular momentum of the
initial data. This tensor is one of the explicit solutions of the
momentum constraint for conformally flat metric studied in \cite{York}.
 Consider the three sphere $S^3$.  
The tensor $\Psi_J^{ab}$ can be extended 
 to an smooth
  tensor on $S^3-\{i_1 \}-\{i_2 \}$, where $i_1$ and $i_2$ are two
  different  points of $S^3$.

  We define $\bar \Psi_J^{ab}$ to be the trace free part with respect
  to $h_{ab}$ of $\Psi_J^{ab}$; and $(\mathcal{L}_h w)^{ab}$ to be the
  conformal Killing operator $\mathcal{L}_h$, with respect to the metric
  $h_{ab}$, acting on a vector $w^a$. We have the following theorem.

\begin{theorem} \label{existencemomentum}
Assume that the metric $h_{ab}$  satisfies (\ref{eq:sobh}) and let $S$ be
$S^3$. Let $i_1$ and $i_2$ two different points in  $S^3$ and  $Q^{ab}\in
W^{1,p}(S)$ an arbitrary symmetric,  trace-free, tensor field. Then,
there exist a unique vector field $w^a\in W^{2,q}(S)$,  $1<q<3/2$, 
such that the 
tensor 
\begin{equation}
  \label{eq:Psiexistence}
  \Psi^{ab}= \bar \Psi_{J}^{ab} +Q^{ab} +(\mathcal{L}_h w)^{ab},
\end{equation}
satisfies the equation $D_a \Psi^{ab}=0$ on $S^3-\{i_1 \}-\{i_2
\}$.
\end{theorem}
The tensor $Q^{ab}$ gives the regular part of the extrinsic curvature, 
which does not contribute to the angular momentum of the initial
data. A  more general 
version of this theorem, which in particular includes linear momentum, 
was proved in \cite{Dain99}. By superposition we obtain solutions
with no trivial angular momentum with three or more asymptotic
ends. It is important to recall that the tensor
$\Psi_{J}^{ab}$ is singular in two points, that is, we need at least
two asymptotic ends. If we insist in having  only one asymptotic end, 
then,  the possible existence of conformal symmetries of the metric
$h_{ab}$ restricts  the allowed value for the angular and
linear momentum of the initial data. For example, if  we chose
$h_{ab}$ to be $h^0_{ab}$, the standard metric of $S^3$, then the
linear and angular momentum of the data have to vanish. For a complete 
discussion of this issue see \cite{Beig96} \cite{Dain99} and also
\cite{Beig00}. 

The linear and angular momentum of the initial data at each end $i_k$
depend only on the tensor $\Psi^{ab}$ and not on the conformal factor
$\theta$, 
they are given by
\begin{equation} \label{momentum}
P^a=-\frac{1}{8\pi} \lim_{\epsilon  \to 0 } \int_{\partial B_{i_k}(\epsilon)} r^2\Psi_{b c} 
(\delta^{b a} -2n^b n^a) n^c dS_\epsilon,
\end{equation}
\begin{equation} \label{angmomentum}
J^a=-\frac{1}{8\pi} \lim_{\epsilon\to 0 } \int_{\partial B_{i_k}(\epsilon)} r\Psi_{c d} \epsilon^{a b c} 
n^d n_b dS_\epsilon,
\end{equation}
where $dS_\epsilon$ is the area element on the two-sphere
$\partial B_{i_k}(\epsilon)$.  If we compute these integrals for the particular tensor given by
(\ref{eq:Psiexistence}) we obtain $P^a=0$ and the angular momentum is
given by the constant vector $J^a$ of (\ref{eq:PsiJp}).

The total mass of the initial data at each end $i_k$ is given by
\begin{equation}
  \label{eq:adm}
  M_{i_k}=\hat  m_{i_k} +2c_{i_k}\sum_{k'=0, \,k'\neq k}^n c_{i_{k'}}\theta_{i_{k'}}(i_k)+2u(i_k), 
\end{equation}
where 
\begin{equation}
  \label{eq:hatm}
\hat  m_{i_k}=2c_{i_k}g_{i_k}(i_k).  
\end{equation}
The constant $\hat  m_{i_k}$ is the mass of the initial data with only 
one end point $i_k$ and $\Psi^{ab}=0$. 
 
Let us mention how some well known initial data are obtained in this
setting. Consider $S^3$ (in the rest of the paper we will always work 
in this manifold) with the standard metric $h^0_{ab}$ and standard
spherical coordinates $(\psi,\vartheta, \phi)$.  The south pole $i_0$
is given by $\psi=0$ and the north pole $i_\pi$ by
$\psi=\pi$. The metric $h^0_{ab}$ has the  form
\begin{equation}
  \label{eq:h0}
  h^0\equiv  d\psi^2 +\sin^2\psi (d\vartheta^2 + \sin^2\vartheta d\phi^2).
\end{equation}
Denote by $(\psi_k,\vartheta_k, \phi_k)$ the coordinates
of the point $i_k$. Using that $h^0_{ab}$ is conformally flat one
easily obtains that the corresponding function $\theta_{i_k}$ is given by
\begin{equation}
  \label{eq:thetas3}
  \theta_{i_k}=\frac{1}{\sqrt{2}(1-\cos \psi\cos \psi_k-\sin \psi\sin
    \psi_k \cos \gamma)^{1/2}}, 
\end{equation}
where
\[
\cos \gamma =\cos \vartheta\cos \vartheta_k+\sin \vartheta\sin
\vartheta_k\cos(\phi-\phi_k).
\]
Note that in this case $\hat m_{i_k}=0$.

Assume that $\Psi^{ab}=0$, this implies $u=0$ by equation
(\ref{gouu}).  If we chose only one end $i_0$, we have
$\theta_0=\theta_{i_0}$ and we obtain the Minkowski initial data.  If
we chose two different ends $i_0$ and $i_1$ we obtain the
Schwarzschild solution, the conformal factor is given by
\begin{equation}
  \label{eq:thetas}
   \theta_s=\theta_{i_0}+ m \theta_{i_1} \sin(\psi_1/2).
\end{equation}
Where $m$ is an arbitrary, positive, constant. We have chosen  the
constants $c_0=1$, $c_1= m \sin(\psi_1/2)$, such that    $m$ is the
total mass  at any of the ends $i_0$ or $i_1$, as we can easily
 verify from equations (\ref{eq:adm})  and  (\ref{eq:thetas3}).

If we have three different points  $i_0$,  $i_1$ and $i_2$, we obtain
the Brill-Lindquist initial data\cite{Brill63}.  The
conformal factor is given by
\begin{equation}
  \label{eq:bl}
  \theta_{ss}=\theta_{i_0}+ m_1 \theta_{i_1} \sin(\psi_1/2)+  m_2
  \theta_{i_2} \sin(\psi_2/2). 
\end{equation}
We have chosen the constants  such that the mass 
at the  at the end $i_0$ is $M_{i_0}=m_1+m_2$. We chose this end point as the place
were the observer is `located'.  
The far limit of the data is  when   $m_1$ or  $m_2$ is equal 
to zero,  or when $\psi_1$ or $\psi_2$ is equal to zero (i.e.; when we
put one of the holes at the `infinity' $i_0$), in both cases  
this data reduce to Schwarzschild data (\ref{eq:thetas}) with mass $m_1$ and
$m_2$ respectively. As a distance parameter $L$ we can chose the
Euclidean distance in $\mathbb{R}^3$ between the points $i_1$ and
$i_2$.  The 
close limit is given by   $i_1=i_2$ (and different from $i_0$), in
this case we obtain  Schwarzschild data with mass $m_0+m_1$. 

Using again that $h^0_{ab}$ is conformally flat and the fact that
(\ref{eq:PsiJp}) is divergence free with respect to the flat metric, we
find that the following tensor
\begin{equation}
  \label{eq:psis3}
  \Psi_0^{ab}=\frac{3}{(\sin \psi)^3}(\hat n^a  \epsilon^{bcd} J_c
  \hat n_d +  
\hat n^b\epsilon^{acd} J_c \hat n_d).
\end{equation}
is trace-free and divergence free with respect to $h^0_{ab}$, where we
have defined $\hat n_a=D_a\psi$.  $\Psi_0^{ab}$ is smooth on
$S^3-\{i_0\}-\{i_\pi\}$ and is singular at the poles.  In an analogous
way we can construct a tensor $\Psi_1^{ab}$ which is singular at two
arbitrary (but different) points $i_0$ and $i_1$. 
Let $\theta_0$ given by
\begin{equation}
  \label{eq:0thetaby1}
  \theta_0= \theta_{i_0}+ m \theta_{i_1} \sin(\psi_1/2).
\end{equation}
Let $u_{by}$ the unique solution of (\ref{gouu}) for   $h^0_{ab}$,
$\Psi_1^{ab}$ and $\theta_0$ given by (\ref{eq:0thetaby1}). The
conformal factor is 
\begin{equation}
  \label{eq:thetaby1}
  \theta_{by}= \theta_0+  u_{by}.
\end{equation}
This is the Bowen-York initial data for one black hole
with spin (the other data discussed there can be obtained a similar
form). The positive constant $m$ is no longer the total mass of the
data.  The mass of the data at the end $i_0$ is given by $m+u_{by}(i_0)$.
The generalization to more than two ends is straightforward: pick up
another tensor $\Psi_2^{ab}$ which is singular at the ends $i_0$ and
$i_2$, define $\Psi_{12}^{ab}=\Psi_1^{ab}+\Psi_2^{ab}$; and consider
$\theta_0$ given by 
\begin{equation}
  \label{eq:0thetaby2}
  \theta_0= \theta_{i_0}+ m_1 \theta_{i_1} \sin(\psi_1/2)+  m_2
  \theta_{i_2} \sin(\psi_2/2).
\end{equation}
Let $\bar u_{by}$ the unique solution of (\ref{gouu}) for $h^0_{ab}$, 
$\Psi_{12}^{ab}$ and (\ref{eq:0thetaby2}). The conformal factor is  given by 
$\theta_{by}= \theta_0+\bar u_{by}$. 
 The mass at $i_0$ is $m_1+m_2+\bar u_{by}(i_0)$. This
data  neither have a far nor a close limit to the Kerr initial data. 

\section{The Kerr initial
  data} \label{Kerrini}
In \cite{Dain00c} it has been proved that the Kerr initial data
satisfies the hypothesis of the existence theorems of section
\ref{constraints}, namely, that the conformal metric satisfies (\ref{eq:posR})
and  (\ref{eq:sobh}), and the conformal extrinsic curvature has a the
form (\ref{eq:Psiexistence}). In this section we briefly summarize
this calculation. 

Consider the Kerr metric in the Boyer-Lindquist
coordinates $(t, \tilde r, \vartheta, \phi )$\cite{Kerr63}
\cite{Boyer67}, with mass
$m$ and angular momentum per mass $a$, such that $m^2>a^2$. We define 
$\delta=\sqrt{m^2-a^2}$.

Take any  slice $t=const$. Denote by $\tilde h^k_{ab}$ the intrinsic three
metric of the slice and by $\tilde \Psi_k^{ab}$ its  extrinsic curvature. 
These  slices are maximal, i. e. 
$\tilde h^k_{ab}\,\tilde \Psi_k^{ab} = 0$.  There exist a coordinate
transformation which maps the region outside the exterior horizon in
to the three
sphere $S^3$ such that $\tilde h^k_{ab}$ is smooth in
$S^3-\{i_0\}-\{i_\pi\}$.  Moreover, there  exist a conformal factor $\theta_k$, which
is singular at the poles  $\{i_0\}$ and $\{i_\pi\}$
\begin{equation}
\label{eq:bThetaK}
\lim_{\psi\rightarrow \pi}(\psi-\pi)\theta_k
=\delta, \quad \lim_{\psi\rightarrow 0}\psi \theta_k=1,
\end{equation}
 such that the rescaled metric 
$h^k_{ab}=\theta_k ^{-4} \tilde h^k_{ab}$ is in the Sobolev space
$W^{4,p}(S^3)$, $p<3$. The conformal metric $h^k_{ab}$ has the form
\begin{equation}
  \label{eq:hk}
h^k_{ab}=h^0_{ab}+a^2  f v_av_b,  
\end{equation}
where   the smooth vector
field 
$v_a$ is given by $v_a\equiv \sin^2\psi \sin^2\vartheta (d\phi)_a$, and the
function $f$, which contains the non-trivial part of the metric, is
given explicity in \cite{Dain00c}.  The only property of this function
that we will use is that it is analytic in the parameters 
 $m$,  $a$  and it satisfies $f\in W^{4,p}(S)$. Since $f$ is
 analytic in $a$,  the Ricci scalar $R$ is also  analytic in $a$.  For
$a=0$ we have that $R=6$, the scalar curvature of $h^0_{ab}$.
Thus, if $a$ is sufficiently small, $R$ will be a positive function on
$S^3$. In the following we will assume the latter condition to be
satisfied.

The metric (\ref{eq:hk}) is axially symmetric, the norm of the Killing vector $\eta^a=(\partial/\partial \phi )^a$ is 
given by
\begin{equation}
  \label{eq:etaKerr}
  \eta^k=\sin^2\psi \sin^2\vartheta(1+a^2f\sin^2\psi \sin^2\vartheta).
\end{equation}

The conformal extrinsic curvature is defined by 
\begin{equation}
 \label{eq:PsiKerrc}
\Psi_k^{ab}=\theta_k^{10} \tilde \Psi_k^{ab}.
\end{equation}
The tensor $\Psi_k^{ab}$ is smooth in  $S^3-\{i_0\}-\{i_\pi\}$ and at the
poles it has the form
\begin{equation}
  \label{eq:PsiKerr}
 \Psi_k^{ab}=\Psi^{ab}_{J}+Q^{ab} 
\end{equation}
where $\Psi^{ab}_{J}$ is given by (\ref{eq:PsiJp}) with $|J|=am$, $Q^{ab}$ is
trace free and $Q^{ab}\in
W^{1,p}(S^3)$.  If $a=0$ then
$\Psi_k^{ab}=0$. Since the Kerr initial data satisfies the constraint, 
we have  that
\begin{equation}
 \label{eq:psik}
D_a \Psi^{ab}_k=0,
\end{equation}
where $D_a$ is the connection of the metric $h^k_{ab}$. 
Using theorems \ref{existenceGreen} and \ref{Beig}, we can decompose
the Kerr conformal factor like
\begin{equation}
  \label{eq:Kerrconf}
  \theta_k=\theta_{i_0}+\delta \theta_{i_\pi} +u_k,
\end{equation}
where the function $u_k\in W^{2,q}(S^3)$.

\section{Close limit initial data with Schwarzschild-like and
  Kerr-like asymptotic ends}\label{sk}

We want to construct  initial data that represent two black-holes,
one of them a Kerr-like  black hole and the other a
Schwarzschild-like black hole. This data will have both a far and a close
limit. The main simplification between this data and the one in the
following section is that the momentum constraint is solved
explicitly, due to the axial symmetry of the conformal metric, i.e.;
we do not make use of theorem \ref{existencemomentum} in this case.

Let $i_1$ an arbitrary point in $S^3$ with coordinates $(\psi_1,\vartheta_1,\phi_1)$, this point will be the location of  the Schwarzschild-like end. Let $m_1$ an arbitrary positive constant. 
Define the following  metric
\begin{equation}
  \label{eq:hsk}
  h^{sk}_{ab}=h^0_{ab}+a^2(\mu  f+ \nu f^K  )v_av_b,
\end{equation}
where $f^K=f(a,m+m_1)$ is the corresponding $f$ function defined by
equation (\ref{eq:hk}) for  a Kerr
initial data with angular momentum per mass $a$ and mass $m+m_1$. 
The functions $\mu$ and $\nu$ will depend smoothly on $m_1$, $m$ and the distance $L$ between  the
point $i_1$ and the point $i_\pi$. These functions have to be chosen in
such a way that    the metric (\ref{eq:hsk}) reduce to  the conformal Kerr
metric $(m,a)$ when $m_1$ is zero or when $L$ goes to infinity (far limit), and
to the  conformal Kerr metric   $(m+m_1, a)$ when $L=0$ (close
limit).  It is clear that the metric (\ref{eq:hsk}) satisfies
(\ref{eq:sobh}) and, for small $a$, (\ref{eq:posR}). 
To ensure that the metric is non-degenerate we impose 
\begin{equation}
  \label{eq:munu+}
 \mu ,\nu \geq 0.
\end{equation}
Define the dimensionless parameter $\epsilon=L^2/(m m_1)$. We impose the
following conditions on $\mu$ and $\nu$. 
In  the far limit 
\begin{equation}
  \label{eq:farsk}
  \epsilon=\infty \Rightarrow  \mu=1, \quad \nu =0;
\end{equation}
and in the close limit
\begin{equation}
  \label{eq:closesk}
  \epsilon=0 \Rightarrow  \mu=0, \quad \nu =1.
\end{equation}
Of course, there is an enormous freedom in the choice of $\mu$ and
$\nu$ which satisfies (\ref{eq:munu+}),
(\ref{eq:farsk}) and (\ref{eq:closesk}). 
For example, we can take
\begin{equation}
  \label{eq:munu}
  \mu=\frac{\epsilon}{1+\epsilon}, \quad  \nu=\frac{1}{1+\epsilon}.
\end{equation}

The metric (\ref{eq:hsk}) is axially symmetric. The norm of the Killing vector is given by
\begin{equation}
 \label{eq:etask}
\eta^{sk}=\sin^2\psi \sin^2\vartheta(1+a^2\sin^2\psi \sin^2\vartheta(\mu  f + \nu  f^K)).
\end{equation}
In appendix \ref{ap} we prove that the following tensor satisfies the
equation $D_a\Psi^{ab}_{sk}=0$ with respect to the metric
(\ref{eq:hsk})
\begin{equation}
  \label{eq:psisk}
  \Psi_{sk}^{ab}=\left(\frac{\eta^{sk}}{\eta^k}\right)^{3/2}\mu \Psi_{k}^{ab}+\left(\frac{\eta^{sk}}{\eta^K}\right)^{3/2}\nu \Psi_{K}^{ab},
\end{equation}
where $\eta^K$  is   the corresponding norm of the
Killing vector for the Kerr metric $(m+m_1,a)$ and $\Psi_{K}^{ab}$ 
is the corresponding conformal extrinsic curvature.  In the far limit
this tensor is equal to $\Psi_{k}^{ab}$ and in the close limit is
equal to
$\Psi_{K}^{ab}$.

Consider the functions $\theta_{i_k}$ corresponding to the metric
$h^{sk}_{ab}$, these functions exist by lemma \ref{existenceGreen}.
Define $\theta_0$ by 
\begin{equation}
  \label{eq:0thetask}
  \theta_0=\theta_{i_0}+(\mu\delta+\nu \delta_K) \theta_{i_\pi} +
  \mu m_1 \sin
  (\psi_1/2) \theta_{i_1},
\end{equation}
where $\delta_K=\sqrt{(m+m_1)^2-a^2}$. Let   $u_{sk}$ be  the unique, non-negative,  solution of
equation (\ref{gouu}), for  the metric (\ref{eq:hsk}), 
extrinsic curvature (\ref{eq:psisk}) and $\theta_0$ given by (\ref{eq:0thetask}).
The solution  exists by
theorem \ref{Beig}.  Then the conformal factor is given by
\begin{equation}
  \label{eq:thetask}
  \theta_{sk}=\theta_0+u_{sk}. 
\end{equation}
Note that the conformal factor has also the proper close and far
limit. This finish the construction. The data has a close limit to a
Kerr initial data with mass $m+m_1$ and angular momentum $(m+m_1)a$. It also 
has far limit: when $m_1=0$ or $L=\infty$ (which implies $\psi_1=0$)
 we obtain exactly the Kerr initial data of mass $m$ and angular
momentum $ma$.  When $m=a=0$ we obtain the Schwarzschild initial data
with mass $m_1$. 

Summarizing, to construct the initial data define $h^{sk}_{ab}$,
$\Psi^{ab}_{sk}$ and $\theta_0$ by equations (\ref{eq:hsk}),
(\ref{eq:psisk}) and (\ref{eq:0thetask}).  Solve equation (\ref{gouu})
for $u_{sk}$ with respect to these tensors.  Define $\theta_{sk}$ by
(\ref{eq:thetask}). Then the physical initial data is given by
$\tilde{h}^{sk}_{ab} = \theta_{sk}^4 h^{sk}_{ab}$ and
$\tilde{\Psi}_{sk}^{ab} = \theta_{sk}^{-10}\Psi_{sk}^{ab}$.

\section{Close limit initial data with two
  Kerr-like asymptotic ends }\label{kk}

Take the Kerr initial data and make a rigid rotation such that the
spin points in to the direction of an arbitrary vector $J_1^a$, and
make a shift of the origin to the coordinate position of an arbitrary
point $i_1$. Let the mass and the modulus of the angular momentum of
this data be $m_1$ and $a_1m_1$.  We obtain a rescaled metric
$h^{k_1}_{ab}=h^0_{ab}+ a^2_1 f_1 v^1_av^1_b$, where $f_1$ and $v^1_a$
are obtained from $f$ and $v_a$ by the rotation and the shift of the
origin, they depend on the coordinates of the point $i_1$ and the
vector $J_1^a$. In an analogous way we define the corresponding
rescaled extrinsic curvature $\Psi_{k_1}^{ab}$. Take another vector
$J_2^a$ and another point $i_2$ and make the same construction.  We
define $f^K_1$ to be the corresponding $f_1$ function of a Kerr
initial data of mass $m_1+m_2$, angular momentum $J_1^a+J_2^a$.  Let
$\Psi^{ab}_{K_1}$ be the corresponding conformal extrinsic curvature.
The function $f_2^K$ and the tensor $\Psi^{ab}_{K_2}$ is defined in
 analogous way, with respect to the origin $i_2$.  Define the
following metric
\begin{equation}
  \label{eq:hkk}
  h_{ab}=h^0_{ab}+\left( a_1^2\mu  f_1 +a_K^2\nu_1f^K_{1}\right)v^1_av^1_b+
\left (a_2^2\mu  f_2 + a_K^2\nu_2f^K_{2}\right)v^2_av^2_b.
\end{equation}
The smooth function $\mu,\nu_1, \nu_2$ will be chosen following
similar arguments as in the previous section, namely 
\begin{equation}
  \label{eq:munukk+}
  \mu,\nu_1,\nu_2\geq 0, 
\end{equation}
we impose the  far limit
\begin{equation}
  \label{eq:farkk}
  \epsilon=\infty \Rightarrow  \mu=1, \quad \nu_1 =\nu_2=0;
\end{equation}
and  the close limit
\begin{equation}
  \label{eq:closeskk}
  \epsilon=0 \Rightarrow \mu=0, \quad \nu_1+\nu_2=1,
\end{equation}
where $\epsilon=L^2/(m_1m_2)$.
In order that this data reduce to the one constructed in the previous
section when  one of the hole is
Schwarzschild,  we impose
\begin{equation}
  \label{eq:farkka}
a_1=0 \Rightarrow  \nu_1=0, \quad   a_2=0 \Rightarrow  \nu_2=0. 
\end{equation}

As in the previous section, it is clear that the metric (\ref{eq:hkk})
satisfies (\ref{eq:sobh}) and (\ref{eq:posR}), for small $a_1$ and
$a_2$.  As an explicit example for the functions $\mu,\nu_1, \nu_2$ we
can take $\mu$ defined by (\ref{eq:munu}) and
\begin{equation}
  \label{eq:munukk}
  \nu_1=\frac{a_1^2}{(a_1^2+a_2^2)(1+\epsilon)},\quad 
  \nu_2=\frac{a_2^2}{(a_1^2+a_2^2)(1+\epsilon)}.
\end{equation}

When the points $i_1$ and $i_2$ are chosen to be on the axis
$\vartheta=0$, the metric (\ref{eq:hkk}) is axially symmetric, we can
apply, as in the previous section, the result of appendix \ref{ap} to
solve for the momentum constraint. But in general, the metric
(\ref{eq:hkk}) is not axially symmetric. To solve the momentum
constraint in this case we use theorem \ref{existencemomentum}.
Define $\Psi^{ab}_{kk}$ by
\begin{equation}
  \label{eq:PsiKK}
  \Psi^{ab}_{kk}=\mu_1\bar \Psi^{ab}_{k_1}+ \nu_1\bar \Psi^{ ab}_{K_1}+
\mu_2\bar \Psi^{ab}_{k_2}+ \nu_2\bar \Psi^{ab}_{K_2}
+(\mathcal{L} w)^{ab},
\end{equation}
where the bar indicate that we take the trace free part with respect
the metric (\ref{eq:hkk}), and  $w^a$ is the unique vector field such
that $ \Psi^{ab}_{kk}$ is divergence free in $\tilde S$ with respect
to the metric (\ref{eq:hkk}). Note that $\Psi^{ab}_{kk}$ has a far
limit to $\Psi^{ab}_{k_1}$ or  $\Psi^{ab}_{k_2}$ and a close limit
($i_1=i_2$) to
$\Psi^{ab}_{K_1}= \Psi^{ab}_{K_2}$. 

The functions $\theta_{i_k}$ for the metric $h^{kk}_{ab}$ exist by
lemma \ref{existenceGreen}, we define $\theta_0$ by 
\begin{equation}
  \label{eq:0ThetaKK}
  \theta_0=\theta_{i_0}+(\mu\delta_1+\nu_1\delta_K) \theta_{i_1}\sin
  (\psi_1/2)+(\mu \delta_2+\nu_2 \delta_K) \theta_{i_2} \sin
  (\psi_2/2). 
\end{equation}
Let $u_{kk}$ be  the unique solution of (\ref{gouu}) with
respect to (\ref{eq:hkk}),  (\ref{eq:ThetaKK}) and (\ref{eq:0ThetaKK}), which  exist
by theorem \ref{Beig}.  The conformal factor is  given by
\begin{equation}
  \label{eq:ThetaKK}
  \theta_{kk}=\theta_0+u_{kk}. 
\end{equation}
We see that this  initial data satisfies the far and close limit with respect
to the Kerr initial data.

\section{Conclusions} \label{fc}
We have constructed a family of initial data that represent two
black-holes and have the desired
far and close limit with respect to the Kerr initial data.   
The main point in the construction is  the choice of the conformal
metric and the conformal extrinsic curvature, given by
(\ref{eq:hkk}) and (\ref{eq:PsiKK}). The conformal metric and  extrinsic curvature  are the `free data', they
determine uniquely a solution of the constraints, provided  they
satisfy certain conditions. These conditions are discussed in the
existence theorems of section \ref{constraints}. 
 In our 
case the data (\ref{eq:hkk}) and (\ref{eq:PsiKK}) satisfy these
conditions since  the Kerr initial data satisfy them, as it was
discussed in section \ref{Kerrini}.  The tensors (\ref{eq:hkk}) and
(\ref{eq:PsiKK}) are, essentially, a  superposition  of three different
Kerr initial data: the two `initial'  Kerr black holes and  the
`final' Kerr black hole. In contrast to the close limit data constructed in
\cite{Price98}, where only one Kerr initial data is used. In that
case,  when one
forces the data to have the topology of two black holes, the equations
produce an unwanted extra singularity.

The functions $\nu$ and
$\mu$, which determined the solution, are only required to satisfy the
far and close limit conditions.  In order to provide an explicit
example, we have presented a particular choice of these functions, but
other choices can be easily constructed. 
For example, in a previous work\cite{Dain00c} we have made the
simpler choice $\nu_1=\nu_2=0$ and $\mu=1$, which do
not have the close limit property.
It
would be very 
 interesting to know  the influence of different choices  in the final form of 
the gravitational waves emitted by the data.

The natural question is whether this construction can be generalized to 
include linear momentum. It is possible to add  an extra term in the extrinsic curvature (\ref{eq:Psiexistence})
which contains the linear momentum of each black hole. The existence
theorem for the momentum constraint  is exactly the same  (see
\cite{Dain99}).  However, we will not 
have either Kerr or Schwarzschild in the far limit,  
 since the
Boyer-Lindquist slices  are
not boosted.  This is  exactly the same situation 
as for the boosted data given in \cite{York}. That is, in order to
produce  initial data which have the far limit to Kerr and also
linear momentum we have to study boosted maximal slices in
Kerr. Even for Schwarzschild this is a non-trivial problem. 
On the other hand, we could impose only a close limit to Kerr. This
can be, in principle,  achieve with a proper choice of the functions
$\mu$ and $\nu$. We will study this generalization in a future work.

\section*{Acknowledgements}
It is a pleasure to thank  J. Baker, S. Husa and C. Lousto for helpful 
discussions and suggestions.

\appendix

\section{Appendix: The momentum constraint in axially symmetric initial data}\label{ap}

We collect here some  results regarding the momentum constraints in
an axially symmetric background which are useful for  constructing 
explicit solutions. What follows is, essentially, a re-writing of the
results  founded in   \cite{Brandt94a} and \cite{Baker99b} in a
coordinate independent way. 

Assume that we have a metric $h_{ab}$ with a  Killing vector $\eta^a$,
which is hypersurface orthogonal. Define the norm $\eta$ by
$\eta=\eta^a \eta^bh_{ab}$. 

Let  $\Psi^{ab}$ the tensor defined by\cite{Hawking73b}
\begin{equation}
  \label{eq:axialpsi}
  \Psi^{ab}=\frac{2}{\eta} S^{(a} \eta^{b)},
\end{equation}
where $S^a$ satisfies
\begin{equation}
 \label{eq:J}
\pounds_\eta S^a=0, \quad S^a\eta_a=0, \quad D_a S^a=0,
\end{equation}
$\pounds_\eta$ is the Lie derivative with respect $\eta$. 

We use the Killing equation $D_{(a}\eta_{b)}=0$,  the fact that
$\eta^a$ is hypersurface orthogonal, (i.e.; it satisfies $D_a
\eta_b=-\eta_{[a}D_{b]} \ln \eta$) and equations (\ref{eq:J})  to
conclude that $\Psi^{ab}$ is trace free and divergence free. 

The metric $h_{ab}$ has the following splitting
\begin{equation}
  \label{eq:sph}
  h_{ab}=e_{ab}+\frac {\eta_a\eta_b}{\eta},
\end{equation}
where $e_{ab}$ is the intrinsic metric of the 2-surface orthogonal to 
$\eta^a$. Let assume that we have another metric $h'_{ab}$ related to $h_{ab}$
by 
\begin{equation}
  \label{eq:h'}
h'_{ab}=e_{ab}+\gamma \frac{\eta_a \eta_b}{\eta}, \quad \pounds_\eta \gamma=0, 
\end{equation}
where $\gamma$ is a positive function. The metric $h'_{ab}$ has a
Killing vector $\eta^a$, with $\eta'_a=\gamma \eta_a$, i.e.; $\gamma=
\eta'/\eta$.  One easily check that the vector ${S'}^a =
S^a/\sqrt{\gamma}$ satisfies equations (\ref{eq:J}) with respect to the
metric $h'_{ab}$. Then, the tensor ${\Psi'}^{ab}$ defined by
\begin{equation}
  \label{eq:axialpsi'}
 \Psi^{'ab}=\frac{1}{\gamma^{3/2}} \Psi^{ab}, 
\end{equation}
is trace free and divergence free with respect to $h'_{ab}$.

The solution of equations  (\ref{eq:J}) can be
written in terms of a scalar potential $\omega$
\begin{equation}
  \label{eq:axialve}
  S^a=\frac{1}{\eta} \epsilon^{abc} \eta_b D_c \omega, \quad
  \pounds_\eta \omega =0. 
\end{equation}
We have restricted ourselves to tensor of the form
(\ref{eq:axialpsi}), since the Kerr extrinsic curvature has precisely
this form. For a more general discussion, and also
for an explicit computation of the potential $\omega$ for Kerr,  see
\cite{Baker99b}.


\end{document}